\documentclass{SCGE}
\usepackage{graphicx,amsmath,amssymb,bm,multirow}
\usepackage{amsthm,mathrsfs}
\usepackage{amsfonts}
\usepackage{dcolumn}
\usepackage{bm}
\usepackage[colorlinks,linkcolor=blue,anchorcolor=blue,citecolor=blue]{hyperref}

\let\citedash\relax
\makeatletter \providecommand{\citedash}{\hbox{-}\penalty\@m}
\makeatother

\newcommand{\beq}{\begin{equation}}
\newcommand{\eeq}{\end{equation}}
\newcommand{\beqn}{\begin{eqnarray}}
\newcommand{\eeqn}{\end{eqnarray}}

\newcommand{\bsub}{ \begin{subequations}}
\newcommand{\esub}{ \end{subequations}}

\renewcommand{\vec}[1]{\mbox{\boldmath $#1$}}

\begin{document}
\begin{picture}(0,0){\rm
\put(0,-20){\makebox[160truemm][l]{\bf {\sanhao\raisebox{2pt}{.}}
Letter  {\sanhao\raisebox{1.5pt}{.}}}}}
\put(0,-34){\jiuwuhao {\textcolor[rgb]{0.5,0.5,0.5}{\sf 
}}}
\end{picture}

\def\bm{\boldsymbol}

\def\dl{\displaystyle}
\def\du{\end{document}}
\def\d{{\rm d}}
\def\e{{\rm e}}
\def\i{{\rm i}}

\Year{2017} %
\Month{April} %
\Vol{XX} 
\No{X} 
\BeginPage{1} 
\AuthorMark{{\rm H. Xia}, et al. Configuration mixing in low-lying spectra of carbon hypernuclei}  
\AuthorMarkCite{H. Xia, H. Mei, and J. M. Yao} 
\DOI{10.1007/s11433-017-9048-2} 
\ArtNo{SCPMA-2017-0140}

\title[Configuration mixing in low-lying spectra of carbon hypernuclei]{Configuration mixing in low-lying spectra of carbon hypernuclei}

\author[1]{XIA HaoJie }{}
\author[2\dag]{MEI Hua}{}
\footnote{\dag Corresponding author (email: meihuayaoyugang@gmail.com)}
\author[2]{YAO JiangMing}{}

\address[{\rm1}]{School of Mechanical and Electrical Engineering, Handan University, Handan  056005, Hebei, China}
\address[{\rm2}]{Department of Physics and Astronomy, University of North Carolina, Chape Hill 27599-3255, North Carolina, USA}

\maketitle \vspace{-3.5mm}{\footnotesize\begin{center} Received April 4, 2017; accepted May 1, 2017; published online May 2, 2017
\end{center}}\vspace*{-5mm}

\begin{center}
\rule{16.5cm}{0.4pt}
\parbox{16.5cm}
{\begin{abstract}
We perform a coupled-channels study of the low-lying states in $^{13,15,17,19}_{~~~~~~~~~~~\Lambda}$C with a covariant energy density functional based microscopic particle-core coupling model. The energy differences of $1/2^-$ and $3/2^-$ states in $^{13}_\Lambda$C and $^{15}_\Lambda$C are predicted to be 0.25 MeV and 0.34 MeV, respectively. We find that configuration mixings in the $1/2^-$ and $3/2^-$ states of $^{15}_\Lambda$C are the weakest among those of $^{13,15,17,19}_{~~~~~~~~~~~\Lambda}$C. It indicates that $^{15}_\Lambda$C provides the best candidate among the carbon hypernuclei to study the  spin-orbit splitting of $p_\Lambda$ hyperon state.
\end{abstract}}

\end{center}\vspace*{-0.6cm}

\begin{center}
\parbox{16.5cm}
{\bf\jiuhao Hypernuclei, Nuclear Density Functional Theory and Extensions, Energy Spectra}
\end{center}

\textwidth=178truemm \textheight=236truemm

\wuhao\vspace*{1.5mm}

\begin{multicols}{2}

\renewcommand{\baselinestretch}{1.08} \baselineskip 12.2pt\parindent=10.8pt

\renewcommand{\thefootnote}


The spectroscopy of hypernuclear low-lying states is very important to understand the structure of hypernuclei and the hyperon impurity effect in atomic nuclei. Several novel phenomena about $\Lambda$ hyperon have already been discovered by studying energy spectra and electromagnetic transitions of low-lying states in $p$-shell hypernuclei. One of them is the shrinkage effect of $\Lambda$ hyperon in $^6$Li, which was indicated from the measurement of $\gamma$-ray transition probability from $5/2^+_1$ to $1/2^+_1$ in $^7_\Lambda$Li. The ratio of $B(E2:5/2^+_1\to 1/2^+_1)$ in $^7_\Lambda$Li to  $B(E2:3^+_1\to 1^+_1)$ in $^6$Li has been converted into the reduction of atomic size by about 19\%~\cite{Tanida01}. Another novel finding is about the weak spin-orbit splitting ($\ell s$) of $p_\Lambda$ state in $^{13}_\Lambda$C. The $\gamma$-rays from the excited $1/2^-_1$ and $3/2^-_1$ states to the ground state were measured following the $^{13}$C($K^-$,$\pi^-$)$^{13}_{~\Lambda}$C reaction. The energy difference between the $1/2^-_1$ and $3/2^-_1$ states was determined to be $152\pm54({\rm stat}) \pm36$({\rm syst})~keV~\cite{Ajimura01}, which has been interpreted as the spin-orbit splitting between 1$p_{1/2}$ and 1$p_{3/2}$ hyperon states in $^{13}_{~\Lambda}$C. Here,  this interpretation relies on the assumption that the $1/2^-_1$ and $3/2^-_1$ states are the pure configuration of $\Lambda p_{1/2}$ and $\Lambda p_{3/2}$ coupled to the ground state ($0^+$) of $^{12}$C~\cite{Auerbach81}, respectively. It is worth mentioning that for most $\Lambda$ hypernuclei, the $1/2^-_1$ and $3/2^-_1$ states cannot be naively interpreted using such a simple picture due to a large effect of configuration mixings~\cite{Motoba85,Mei2014,Mei2015,Mei2016}. It has been found in a recent study that the mixing amplitude is negligibly small in spherical and weakly-deformed hypernuclei, but strongly increases as the core nucleus undergoes a transition to a well-deformed shape~\cite{Mei17-Sm}. This perturbs the interpretation of their energy difference as the spin-orbit splitting for the $p_\Lambda$ state.

Considering the great success of covariant density functional theory for atomic nuclei~\cite{Meng2006470} and hypernuclei~\cite{Hagino_Rev}, in this letter, we examine the configuration mixing in the low-lying states of $^{13,15,17,19}_{~~~~~~~~~~~\Lambda}$C with a novel microscopic particle-core coupling model built on a multi-reference covariant density functional theory (MR-CDFT). The MR-CDFT has been successfully adopted to describe the low-lying states of carbon isotopes~\cite{Yao11-C}. It provides a good starting point for studying the hypernuclear low-lying states. Special emphasis will be placed upon the energies and the ingredients of the wave functions for the lowest $1/2^-$ and $3/2^-$ states in the carbon isotopes.


 In the microscopic particle-core coupling model~\cite{Mei2014,Mei2015,Mei2016}, wave functions of the low-lying states of $\Lambda$ hypernuclei  are constructed as
\begin{equation}
 \label{wavefunction}
 \displaystyle \Psi_{JM}(\vec{r},\{\vec{r}_i\})
 =\sum_{n,\ell, j, I}  {\mathscr R}_{j\ell n I}(r) [{\mathscr Y}_{j\ell }(\hat{\vec{r}})\otimes
\Phi_{n I}(\{\vec{r}_i\})]^{(JM)},
\end{equation}
where $\vec{r}$ and $\vec{r}_i$ are the coordinates of the $\Lambda$ hyperon and the
nucleons, respectively. ${\mathscr R}_{j\ell n I}(r)$ and ${\mathscr Y}_{j\ell}(\hat{\vec{r}})$ are the radial wave function and the spin-angular wave function for the $\Lambda$-particle, respectively. The index $n = 1, 2, \ldots$ distinguishes different core states for a given angular momentum $I$.  The nuclear core states $\vert\Phi_{nI}\rangle$ are determined from the MR-CDFT~\cite{Yao09,Yao10,Yao11-Mg,Yao14},
\begin{equation}
  \label{GCMwf}
 \vert\Phi_{nI}\rangle=\sum_{\beta}  F_{nI}(\beta)\vert\Phi_{IM,NZ}(\beta)\rangle.
 \end{equation}
The symmetry-conserving reference states are a set of axially deformed mean-field states
$\vert \beta\rangle$ projected onto  angular momentum ($I$) and particle numbers $(N, Z)$
\beq
\label{MF}
 \vert\Phi_{IM,NZ}(\beta)\rangle=\hat{P}^I_{MK=0}\hat{P}^N \hat{P}^Z \vert \beta \rangle
\eeq
where the $\hat{P}^I_{MK}$, and $\hat{P}^N, \hat{P}^Z$ are projection operators.
The mean-field states $\vert \beta\rangle$ are obtained by  relativistic mean-field plus BCS calculations with a constraint on the quadrupole deformation parameter $\beta$. The weight function $F_{nI}(\beta)$ and the energy $E_{nI}$ of the state $\vert\Phi_{nI}\rangle$ are the solutions of the Hill-Wheeler-Griffin (HWG) equation~\cite{Ring80}. The Hamiltonian kernel is calculated with the mixed-density prescription, i.e., the off-diagonal elements of the energy overlap (sandwiched by two different reference states) take the same energy functional form as that of the diagonal one provided that all the densities and currents are replaced by the mixed ones~\cite{Yao09,Yao10}.

The Hamiltonian $\hat H$ for the whole $\Lambda$ hypernucleus is chosen as~\cite{Mei2016}
\begin{equation}
\hat H =\hat T_\Lambda +  \hat H_{\rm c}+ \sum^{A_c}_{i=1} \hat{V}^{N\Lambda}(\vec{r},\vec{r}_{i}),
\label{eq:H}
\end{equation}
where $\hat T_\Lambda$ is the kinetic energy of the hyperon, and $\hat H_c$ is the many-body Hamiltonian for the core nucleus, satisfying $\hat H_c \vert\Phi_{n I}\rangle= E_{n I} \vert\Phi_{n I}\rangle$. The third term on the right side of Eq. (\ref{eq:H}) represents the interaction term between the $\Lambda$ particle and the nucleons in the core nucleus, where $A_c$ is the mass number of the core nucleus. The nucleon-$\Lambda$ ($N\Lambda$) interaction takes the form derived from a relativistic point-coupling energy functional~\cite{Tanimura2012}
\begin{align}
\label{Scalar}
\hat{V}^{N\Lambda}_{\rm S}(\vec{r},\vec{r}_i)=& \alpha_S^{N\Lambda} \gamma^0_\Lambda
\delta(\vec{r}-\vec{r}_i)\gamma^0_N
+\delta_S^{N\Lambda}\gamma^0_\Lambda
\Big[\overleftarrow{\nabla}^2 \delta(\vec{r}-\vec{r}_i) \nonumber \\
+& \delta(\vec{r}-\vec{r}_i)\overrightarrow{\nabla}^2+ 2 \overleftarrow{\nabla}
\cdot\delta(\vec{r}-\vec{r}_i)
 \overrightarrow{\nabla}\Big]\gamma^0_N,\\
 \label{Vector}
\hat{V}^{N\Lambda}_{\rm V}(\vec{r},\vec{r}_i)=&\alpha_V^{N\Lambda} \delta(\vec{r}-\vec{r}_i)
+\delta_V^{N\Lambda} \Big[\overleftarrow{\nabla}^2 \delta(\vec{r}-\vec{r}_i)
 \nonumber \\
+& \delta(\vec{r}-\vec{r}_i) \overrightarrow{\nabla}^2+ 2
\overleftarrow{\nabla}\cdot\delta(\vec{r}-\vec{r}_i) \overrightarrow{\nabla}\Big],\\
 \label{Tensor}
\hat{V}^{N\Lambda}_{\rm Ten}(\vec{r},\vec{r}_i)=&
i\alpha_T^{N\Lambda}\gamma^0_\Lambda\Big[\overleftarrow{\nabla} \delta(\vec{r}-\vec{r}_i)
+\delta(\vec{r}-\vec{r}_i)\overrightarrow{\nabla}\Big]\cdot \vec{\alpha}.
\end{align}

The equation $\hat{H}|\Psi_{JM}\rangle=E_J|\Psi_{JM}\rangle$ is transformed into coupled-channels equations in relativistic framework. Therefore, the ${\mathscr R}_{j\ell n I}(r)$ is a four-component wave function that
is solved by expanding on a set of radial wave functions of spherical harmonic oscillator. The expansion coefficients are determined by solving the coupled-channel equations, in which all the potentials coupling different channels are related to transition densities between low-lying states of core nuclei from the MR-CDFT calculation. More details about the microscopic-particle core coupling model can be found in Refs.~\cite{Mei2014,Mei2015,Mei2016}.


 The wave functions of the mean-field states $\vert \beta\rangle$ in Eq.(\ref{MF}) are products of single-particle wave functions which are obtained by solving the Dirac equation self-consistently on a harmonic oscillator basis with 12 major shells. The parameter sets PC-F1~\cite{BurRMFPC02} and PCY-S4~\cite{Tanimura2012} are adopted for the effective nucleon-nucleon and $N\Lambda$ interaction, respectively. In the particle-number and angular-momentum projection calculation, the number of mesh points for gauge angle in $[0,\pi]$ is chosen to be 9 and that for Euler angle $\theta$ in the interval $[0,\pi]$ is chosen to be 16. The radial wave function in Eq.(\ref{wavefunction}) is expanded on the radial part of spherical harmonic oscillator basis with 18 major shells.

\begin{figure}[h]
  \centering\includegraphics[width=10cm]{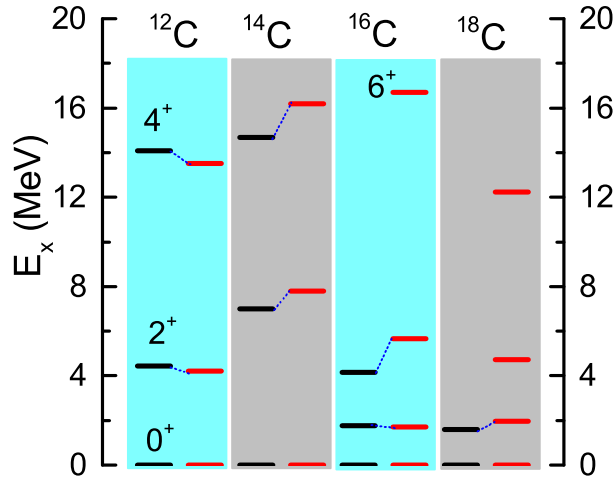}
  \centering\caption{(Color online) The low-lying energy spectra of $^{12,14,16,18}$C from the multi-reference covariant density functional calculation (right), in comparison with data (left)~\cite{NNDC}. }
  \label{Spectrum1}
\end{figure}

 Figure~\ref{Spectrum1} shows the low-lying states of $^{12,14,16,18}$C from the MR-CDFT calculation, in comparison with data~\cite{NNDC}. Since the exact particle-number projection is adopted in the present calculation, the predicted energy spectra are somewhat different from those presented in Ref.~\cite{Yao11-C}, where only an approximate way~\cite{Yao11-Mg} was adopted to take care of the particle-number conserving effect. One can see that the main characters of the energy spectra are reproduced rather well. Our calculation gives a vibrational-like spectrum for $^{14}$C and rotational-like spectra for other isotopes.

 Table~\ref{tab:table1} lists the binding energies $B_\Lambda$ of $\Lambda$ hyperon for the ground states of $^{A_c+1}_{~~~~~\Lambda}$C from the microscopic particle-core coupling model calculation. The $\Lambda$ binding energy is calculated as the energy difference between the $1/2^+_1$ state in $^{A_c+1}_{~~~~~\Lambda}$C and the $0^+_1$ state in $^{A_c}$C. It is shown that the calculated $B_\Lambda$ is 13.22 MeV for $^{13}_\Lambda$C (compared to the data $11.38(5)$ MeV~\cite{Hashimoto06}) and increases globally with neutron number, except for $^{15}_{~\Lambda}$C. For comparison, the intrinsic quadrupole deformation parameters for the hypernuclei $^{A_c+1}_{~~~~~\Lambda}$C (with the $\Lambda$ on the lowest-energy state) and core nuclei $^{A_c}$C from pure mean-field calculations are also provided in Table~\ref{tab:table1}. One can see that the quadrupole deformation of the whole $\Lambda$ hypernuclei is decreased in comparison with that of core nuclei. This finding has already been found in many studies~\cite{Zhou07,Win08,Win11,Lu2011,Lu14,Xue2015}.

\begin{tablehere}
\caption{The binding energies (in unit of MeV) of $\Lambda$ hyperon in ground states of carbon isotopes from the microscopic particle-core coupling model calculation. The intrinsic quadrupole deformation parameters $\beta$ and $\beta_c$ from the pure mean-field calculations for hypernuclei and nuclear core are also given.}
\label{tab:table1}
\vspace{-1mm}\footnotesize
\begin{center} \doublerulesep 0.1pt \tabcolsep 8pt
\begin{tabular}{lccccc}
\hline
 & $^{13}_{~\Lambda}$C~~~~~~ &  $^{15}_{~\Lambda}$C~~~~~~ &$^{17}_{~\Lambda}$C~~~~~~ & $^{19}_{~\Lambda}$C~~~~~~ \\
 \hline
B$_{\Lambda}$        & 13.22~~~~~~~~~ &   13.04~~~~~~~~~    &  13.52~~~~~~~~~    & 14.20~~~~~~~~~ \\
$\beta$ ($\beta_c$)  & 0.00~~(0.00)   &  0.00~~(0.00) &    0.27~~(0.34) &  0.34~~(0.41)\\
 \hline
\end{tabular}
\end{center}
\end{tablehere}

\begin{figure}[ht]
  \centering\includegraphics[width=14cm]{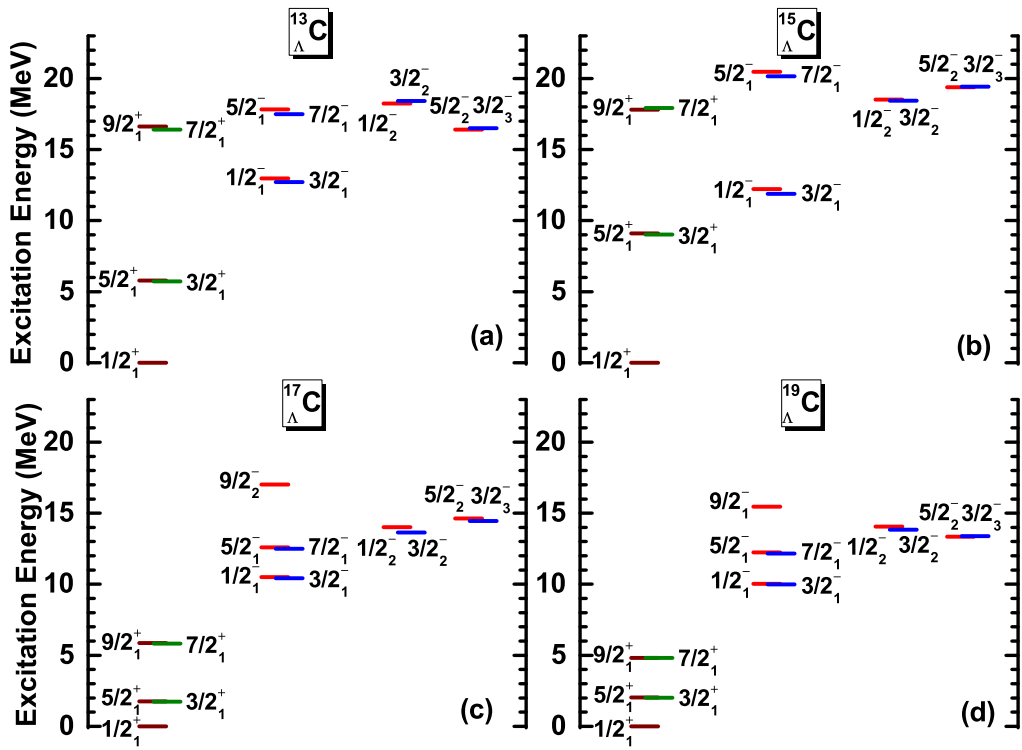}
  \centering\caption{(Color online) The low-lying energy spectra of $^{13,15,17,19}_{~~~~~~~~~~~~\Lambda}$C from the microscopic particle-core coupling model. }
  \label{Spectrum2}
\end{figure}

Figure~\ref{Spectrum2} displays the low-lying energy spectra of $^{13,15,17,19}_{~~~~~~~~~~~~\Lambda}$C. The positive-parity ground-state band with spin-parity of $(I\pm1/2)^+$ in all the four carbon hypernuclei shares a similar structure to that for the corresponding core nucleus and their wave functions are dominated by the configuration of $[\Lambda s_{1/2}\otimes I^+]$, where $\Lambda s_{1/2}$ denotes the $\Lambda$ particle in the $s_{1/2}$ state. In contrast, the low-lying negative-parity states $J^-$ show an admixture of the $[\Lambda p_{1/2}\otimes I^+]$ and the $[\Lambda p_{3/2}\otimes (I\pm2)^+]$ configurations, as shown in Table~\ref{table:Component}. One can see that the $1/2^-_1$ state in $^{13}_\Lambda$C is also an admixture of $[\Lambda p_{1/2}\otimes 0^+_1]$ and $[\Lambda p_{3/2}\otimes 2^+_1]$ with the mixing amplitude of $0.885$ and $0.105$, while the  mixing amplitude for $3/2^-_1$ in  $^{13}_\Lambda$C is  $0.045$ and $0.921$, respectively. It indicates that the interpretation of the energy difference between the $1/2^-_1$ and $3/2^-_1$ states $^{13}_\Lambda$ as the splitting of single-$p_\Lambda$ states~\cite{Auerbach81,Tanida01} is questionable.  In contrast, the mixing amplitude in $^{15}_\Lambda$C is weaker than that in $^{13}_\Lambda$C  and thus $^{15}_\Lambda$C provides a more ideal hypernucleus with which to extract the $\ell s$ splitting of $p_\Lambda$ state.
The stronger configuration-mixing amplitude in $^{13}_\Lambda$C than that in $^{15}_\Lambda$C can be traced back to the relatively larger quadrupole collectivity of corresponding core nucleus. Previous study~\cite{Yao11-C} has shown that the energy surface of $^{12}$C as a function of deformation $\beta$  is much softer than that of the closed-shell nucleus $^{14}$C and thus $^{12}$C possesses a larger quadrupole collectivity from dynamic shape fluctuation.

Table~\ref{tab:splitting} lists the energies of the  $1/2^-_1$ and $3/2^-_1$ states  in carbon hypernuclei and their energy differences $\Delta E$. We note that the $1/2^-_1$ state in all the four hypernuclei is predicted to be higher than the $3/2^-_1$ state. For $^{13}_\Lambda$C, the predicted  $\Delta E$  is 0.253  MeV, close to the data $\Delta E=0.152(90)$ MeV~\cite{Hashimoto06}. For $^{15}_\Lambda$C, the energy difference of the $1/2^-_1$ and $3/2^-_1$ states is predicted to be 0.344 MeV, about 0.1 MeV larger than that in $^{13}_\Lambda$C. In contrast to the cases in $^{13,15}_{~~~~~\Lambda}$C, this value is only 67 keV and 33 keV in $^{17,19}_{~~~~~\Lambda}$C, respectively. One can see from Table~\ref{table:Component} that the energy difference of the $1/2^-_1$ and $3/2^-_1$ states in $^{17,19}_{~~~~~\Lambda}$C cannot be interpreted as the spin-orbit splitting of the $p_\Lambda$ state due to the large configuration mixing. As found in the recent study for Sm hypernuclei~\cite{Mei17-Sm}, the energy difference of the $1/2^-_1$ and $3/2^-_1$ states is monotonically decreasing as the amplitude of configuration mixing increases. A similar phenomenon is also observed in the carbon hypernuclei. In short, our results indicate that $^{15}_\Lambda$C is a more ideal hypernucleus than $^{13}_\Lambda$C to extract the $\ell s$ splitting of the $p_\Lambda$ state.

\begin{tablehere}
 \caption{The probability $P$ (defined as  $P\equiv\int dr r^2 \vert {\mathscr R}_{j\ell n I}(r)\vert^2$) of the dominant components in the wave functions for some selected negative-parity states. The components with   probabilities
smaller than 0.001 are not given.}
 \label{table:Component}
\vspace{-1mm} \footnotesize
\begin{center} \doublerulesep 0.1pt \tabcolsep 7pt
\begin{tabular}{cc|cccc}
  \hline\hline
$J^\pi$    &$(lj)\otimes I^\pi_{n}$&$^{13}_{~\Lambda}$C &$^{15}_{~\Lambda}$C &$^{17}_{~\Lambda}$C &$^{19}_{~\Lambda}$C
\\     \hline
$1/2^-_1 $  &$p_{1/2}\otimes0_1^+$ &$0.8853$ &$0.9643$ &$0.5477$ &$0.5705$  \\
  $$        &$p_{3/2}\otimes2_1^+$ &$0.1054$ &$0.0339$ &$ 0.4400$ &$0.4066$  \\

$3/2^-_1 $  &$p_{1/2}\otimes2_1^+$ &$0.0453$ &$0.0148$ &$0.1913$ &$0.1865$      \\
            &$p_{3/2}\otimes0_1^+$ &$0.9207$ &$0.9738$ &$0.6071$ &$0.6184$  \\
            &$p_{3/2}\otimes2_1^+$ &$0.0243$ &$0.0089$ &$0.1897$ &$0.1682$     \\

$5/2^-_1 $  &$p_{1/2}\otimes2_1^+$ &$0.7643$ &$0.7916$ &$0.5823$ &$0.5230$      \\
            &$p_{3/2}\otimes4_1^+$ &$0.0368$ &$0.0107$ &$0.1632$ &$0.2766$  \\
            &$p_{3/2}\otimes2_1^+$ &$0.1905$ &$0.1491$ &$0.2415$ &$0.1770$      \\

$7/2^-_1 $  &$p_{1/2}\otimes4_1^+$ &$0.0216$ &$0.0060$ &$0.0883$ &$0.1506$          \\
            &$p_{3/2}\otimes2_1^+$ &$0.9613$ &$0.9603$ &$0.8466$ &$0.7317$  \\
            &$p_{3/2}\otimes4_1^+$ &$$ &$$ &$0.0535$ &$0.0979$     \\
 \hline

 $1/2^-_2 $ &$p_{1/2}\otimes0_1^+$ &$0.4547$ &$0.9351$ &$0.4292$ &$0.2969$ \\
            &$p_{3/2}\otimes2_1^+$ &$0.5235$ &$0.0645$ &$0.5306$ &$0.2864$ \\

$3/2^-_2 $   &$p_{1/2}\otimes2_1^+$ &$0.2343$ &$0.0356$  &$0.1257$  &$0.3076$  \\
             &$p_{3/2}\otimes2_1^+$ &$0.2043$ &$0.0063$  &$0.4830$  &$0.0580$  \\
             &$p_{3/2}\otimes0_1^+$ &$0.5272$ &$0.9575$  &$0.3513$  &$0.2731$ \\
 $5/2^-_2 $ &$p_{1/2}\otimes2_1^+$  &$0.1955$ &$0.1614$  &$0.2370$  &$0.1473$ \\
            &$p_{3/2}\otimes2_1^+$  &$0.7943$ &$0.8338$  &$0.6827$  &$0.5479$ \\
             \hline
$3/2^-_3 $ &$p_{1/2}\otimes2_1^+$ &$0.4201$ &$0.3841$ &$0.6045$ &$0.1511$ \\
           &$p_{3/2}\otimes2_1^+$ &$0.5665$ &$0.6078$ &$0.2698$ &$0.5138$ \\

\hline\hline
 \end{tabular}
\end{center}
\end{tablehere}
\end{multicols}

\begin{multicols}{2}

\begin{tablehere}
\caption{The excitation energies $E_x$ (in unit of MeV) of the lowest $1/2^-$ and $3/2^-$ states  and their difference $\Delta E$ in carbon hypernuclei.}
\label{tab:splitting}
\vspace{-1mm}\footnotesize
\begin{center} \doublerulesep 0.1pt \tabcolsep 12pt
\begin{tabular}{lccccc}
\hline
 & $^{13}_{~\Lambda}$C &  $^{15}_{~\Lambda}$C &$^{17}_{~\Lambda}$C & $^{19}_{~\Lambda}$C \\
 \hline
   $E_x(1/2^-)$   &   12.964       &  12.224  &   10.498    & 10.027 \\
   $E_x(3/2^-)$   &   12.711       &  11.880   &   10.431   &  9.994 \\
\hline
$\Delta E$       &    0.253   & 0.344 &	0.067 &	0.033\\
 \hline
\end{tabular}
\end{center}
\end{tablehere}


We have investigated the low-lying states in $^{13,15,17,19}_{~~~~~~~~~~~\Lambda}$C with a microscopic particle-core coupling model based on the MR-CDFT. It has been found that the positive-parity ground-state band  of hypernuclei with spin-parity of $J^\pi=(I\pm1/2)^+$ is dominated by the configuration of $[\Lambda s_{1/2}\otimes I^+]$ and shares a similar structure to that for the core nucleus with spin-parity $I^+$. In contrast, the low-lying negative-parity states $J^-$ are admixtures of the $[\Lambda p_{1/2}\otimes I^+]$ and the $[\Lambda p_{3/2}\otimes (I\pm2)^+]$ configurations. Among all the four carbon hypernuclei, $^{15}_\Lambda$C stands out as the best candidate to extract the spin-orbit splitting of the $p_\Lambda$ state because about 97\% of the wave functions for the $1/2^-_1$ and $3/2^-_1$ states are the configurations of $[\Lambda p_{1/2}\otimes 0^+_1]$ and $[\Lambda p_{3/2}\otimes 0^+_1]$, respectively. Their energy difference turns out to be 0.344 MeV.  We have found that the $1/2^-_1$ and $3/2^-_1$ states in $^{13}_\Lambda$C have a slightly larger configuration mixing than those in $^{15}_\Lambda$C. From this point of view, we conjecture that the previously measured energy difference between the $1/2^-_1$ and $3/2^-_1$ states in $^{13}_\Lambda$C~\cite{Tanida01} underestimates the  $\ell s$ splitting of single-$\Lambda_p$ states.  Therefore, a new measurement on hypernuclear $\gamma$-ray spectroscopy for $^{15}_\Lambda$C is suggested to confirm our conclusions. This kind of measurement is feasible with the Japan Proton Accelerator Research Complex (J-PARC) facility~\cite{Hashimoto06}.

\vspace*{2mm} \Acknowledgements{\bahao This work was supported in part by the National Natural Science Foundation of China under Grant Nos. 11575148, and 11305134.}

\end{multicols}


\begin{thebibliography}{90}

\bibitem{Tanida01}
  K.~Tanida, H.~Tamura, D.~Abe, H.~Akikawa, K.~Araki, H.~Bhang, T.~Endo,
  Y.~Fujii, T.~Fukuda, O.~Hashimoto, K.~Imai, H.~Hotchi, Y.~Kakiguchi, J.~H.
  Kim, Y.~D. Kim, T.~Miyoshi, T.~Murakami, T.~Nagae, H.~Noumi, H.~Outa,
  K.~Ozawa, T.~Saito, J.~Sasao, Y.~Sato, S.~Satoh, R.~I. Sawafta, M.~Sekimoto,
  T.~Takahashi, L.~Tang, H.~H. Xia, S.~H. Zhou, and L.~H. Zhu,
\newblock {\em Phys. Rev. Lett.} 86, 1982 (2001).

\bibitem{Ajimura01}
S.~Ajimura, H.~Hayakawa, T.~Kishimoto, H.~Kohri, K.~Matsuoka, S.~Minami,
  T.~Mori, K.~Morikubo, E.~Saji, A.~Sakaguchi, Y.~Shimizu, M.~Sumihama, R.~E.
  Chrien, M.~May, P.~Pile, A.~Rusek, R.~Sutter, P.~Eugenio, G.~Franklin,
  P.~Khaustov, K.~Paschke, B.~P. Quinn, R.~A. Schumacher, J.~Franz, T.~Fukuda,
  H.~Noumi, H.~Outa, L.~Gan, L.~Tang, L.~Yuan, H.~Tamura, J.~Nakano,
  T.~Tamagawa, K.~Tanida, and R.~Sawafta,
\newblock {\em Phys. Rev. Lett.} 86, 4255 (2001).


\bibitem{Auerbach81}
E.~H. Auerbach, A.~J. Baltz, C.~B. Dover, A.~Gal, S.~H. Kahana, L.~Ludeking,
  and D.~J. Millener,
\newblock {\em Phys. Rev. Lett.} 47, 1110 (1981).

\bibitem{Motoba85}
T.~Motoba, H.~Band\=o, K.~Ikeda, and T.~Yamada,
\newblock {\em Prog. Theor. Phys. Suppl.} 81, 42 (1985).


\bibitem{Mei2014}
H.~Mei, K.~Hagino, J.~M. Yao, and T.~Motoba,
\newblock {\em Phys. Rev. C} 90, 064302 (2014).

\bibitem{Mei2015}
H.~Mei, K.~Hagino, J.~M. Yao, and T.~Motoba,
\newblock Microscopic study of low-lying spectra of
\newblock {\em Phys. Rev. C} 91, 064305 (2015).

\bibitem{Mei2016}
H.~Mei, K.~Hagino, J.~M. Yao, and T.~Motoba,
\newblock {\em Phys. Rev. C} 93, 044307 (2016).

\bibitem{Mei17-Sm}
H.~Mei, K.~Hagino, J.~M. Yao, and T.~Motoba,
\newblock {\em arXiv preprint arXiv:1704.02258}, 2017.


\bibitem{Meng2006470}
 J. Meng, H.~Toki, S.G. Zhou, S.Q. Zhang, W.H. Long, and L.S. Geng,
\newblock {\em Prog. Part. Nucl. Phys.} 57, 470 (2006).

  \bibitem{Hagino_Rev}
K.~Hagino and J.~M. Yao,  {\em International Review of Nuclear Physics}, Vol. 10, edited by J. Meng (World Scientific, Singapore, 2016), pp. 263-303.

\bibitem{Yao11-C}
J.~M. Yao, J.~Meng, P.~Ring, Z.~X. Li, Z.~P. Li, and K.~Hagino,
\newblock {\em Phys. Rev. C} 84, 024306 (2011).



\bibitem{Yao14}
J.~M. Yao, K.~Hagino, Z.~P. Li, J.~Meng, and P.~Ring,
\newblock {\em Phys. Rev. C} 89, 054306 (2014).

\bibitem{Yao11-Mg}
J.~M. Yao, H.~Mei, H.~Chen, J.~Meng, P.~Ring, and D.~Vretenar,
\newblock {\em Phys. Rev. C} 83, 014308 (2011).

\bibitem{Yao09}
J.~M. Yao, J.~Meng, P.~Ring, and D.~Pena Arteaga,
\newblock {\em Phys. Rev. C} 79, 044312 (2009).


\bibitem{Yao10}
J.~M. Yao, J.~Meng, P.~Ring, and D.~Vretenar,
\newblock {\em Phys. Rev. C} 81, 044311 (2010).


\bibitem{Ring80}
P.~Ring and P.~Schuck,
\newblock {\em The Nuclear Many-Body Problem}, (Springer-Verlag, New York,
  1980).


\bibitem{Tanimura2012}
Y.~Tanimura and K.~Hagino,
\newblock {\em Phys. Rev. C} 85, 014306 (2012).


\bibitem{BurRMFPC02}
T.~B\"urvenich, D.~G. Madland, J.~A. Maruhn, and P.-G. Reinhard,
\newblock {\em Phys. Rev. C} 65, 044308 (2002).


\bibitem{NNDC}
National Nuclear Data Center,
\newblock {\em [http://www.nndc.bnl.gov/].}



\bibitem{Hashimoto06}
O.~Hashimoto and H.~Tamura,
\newblock {\em Prog. Part. Nucl. Phys.} 57, 564 (2006).

\bibitem{Zhou07}
X.-R. Zhou, H.-J. Schulze, H.~Sagawa, C. X. Wu, and E.-G. Zhao,
\newblock {\em Phys. Rev. C} 76, 034312 (2007).


\bibitem{Win08}
Myaing~Thi Win and K.~Hagino,
\newblock {\em Phys. Rev. C} 78, 054311 (2008).


\bibitem{Lu14}
B.-N. Lu, E. Hiyama, H. Sagawa, and S.-G. Zhou,
\newblock {\em Phys. Rev. C} 89, 044307 (2014).

\bibitem{Win11}
Myaing~Thi Win, K.~Hagino, and T.~Koike,
\newblock {\em Phys. Rev. C} 83, 014301 (2011).

\bibitem{Lu2011}
B.-N. Lu, E.-G Zhao, and S.-G. Zhou,
\newblock {\em Phys. Rev. C} 84, 014328 (2011).


\bibitem{Xue2015}
W.~X. Xue, J.~M. Yao, K.~Hagino, Z.~P. Li, H.~Mei, and Y.~Tanimura,
\newblock {\em Phys. Rev. C} 91, 024327 (2015).

\end{thebibliography}
\end{document}